\documentclass[
showpacs,
floatfix,
aps,
prl,
twocolumn,
superscriptaddress,
amssymb,
]{revtex4-1}

\usepackage{amssymb}
\usepackage{latexsym}
\usepackage[dvips]{graphicx}
\usepackage{amsmath}

\usepackage{graphicx}
\usepackage{dcolumn}
\usepackage{amsfonts}
\usepackage{bm}
\usepackage{epsfig}

\newcommand{\be}{\begin{equation}}
\newcommand{\ee}{\end{equation}}
\newcommand{\bea}{\begin{eqnarray}}
\newcommand{\eea}{\end{eqnarray}}

\def\bip{B_\parallel}

\def\A{A_{\theta}}

\def\rxx{R_{xx}}
\def\ryy{R_{yy}}

\newcommand{\rfig}[1]{Fig.\,\ref{#1}}

\newcommand{\rref}[1]{Ref.\,\onlinecite{#1}}

\begin{document}
\title{Anisotropic Transport in Ge/SiGe Quantum Wells in Tilted Magnetic Fields}
\author{Q.~Shi}
\affiliation{School of Physics and Astronomy, University of Minnesota, Minneapolis, Minnesota 55455, USA}
\author{M.~A.~Zudov}
\email[Corresponding author: ]{zudov@physics.umn.edu}
\affiliation{School of Physics and Astronomy, University of Minnesota, Minneapolis, Minnesota 55455, USA}
\author{C.~Morrison}
\affiliation{Department of Physics, University of Warwick, Coventry, CV4 7AL, United Kingdom}
\author{M.~Myronov}
\affiliation{Department of Physics, University of Warwick, Coventry, CV4 7AL, United Kingdom}

\begin{abstract}
We report on a strong transport anisotropy in a 2D hole gas in a Ge/SiGe quantum well, which emerges only when both perpendicular and in-plane magnetic fields are present.
The ratio of resistances, measured along and perpendicular to the in-plane field, can exceed $3\times 10^4$.
The anisotropy occurs in a wide range of filling factors where it is determined {\em primarily} by the tilt angle.
The lack of significant anisotropy without an in-plane field, easy tunability, and persistence to higher temperatures and filling factors set this anisotropy apart from nematic phases in GaAs/AlGaAs.
\end{abstract}
\maketitle

Strong transport anisotropies were experimentally discovered in a high-mobility 2D electron gas (2DEG) in GaAs/AlGaAs heterostructures subject to strong perpendicular magnetic fields and low temperatures ($T < 0.1$ K) \cite{lilly:1999a,du:1999}.
This remarkable phenomenon is marked by the resistivity minima (maxima) in the easy (hard) transport direction near half-integer filling factors, $\nu = 2N + 1 \pm 1/2$ ($2 \leq N \leq 6$), where $N$ is the Landau level index.
The effect has been interpreted in terms of ``stripes"
\citep{koulakov:1996,fogler:1996, rezayi:1999}, or a nematic phase \cite{fradkin:1999,fradkin:2000,fradkin:2010}, formed due to interplay between exchange and direct Coulomb interactions.
The origin of the native anisotropy, i.e., how its axes are chosen, is still being debated \cite{sodemann:2013,kovudayur:2011}.

It is well known that an in-plane magnetic field $\bip$ applied along the easy direction usually switches the anisotropy axes \cite{lilly:1999b,pan:1999,jungwirth:1999,stanescu:2000}, aligning the hard axis parallel to $\bip$.
Applying $\bip$ along the hard axis could either increase or decrease the anisotropy \cite{lilly:1999b,cooper:2002,pan:1999} and, sometimes, also switch easy and hard axes \cite{lilly:1999b}. 
In addition, $\bip$ can induce anisotropy in isotropic states, such as fractional quantum Hall (QH) states at $\nu=5/2,\,7/2$ \cite{lilly:1999b,pan:1999} and $\nu=7/3$ \cite{xia:2011}. 
When $\bip$ is applied, these states either become anisotropic compressible states, but, occasionally, the anisotropy coexists with the QH effect \cite{xia:2011,mulligan:2011,liu:2013b}.
The effect of $\bip$ can also depend on its orientation with respect to the crystallographic axes, even when the initial state is isotropic \cite{zhang:2010,zhang:2012}.

Another class of $\bip$-induced anisotropies appears at \emph{integer} $\nu$, when two Landau levels are brought into coincidence \cite{zeitler:2001,pan:2001,luhman:2006}.
For example, \rref{zeitler:2001} reported a strong anisotropy at $\nu = 4$ of a 2DEG in Si/SiGe in a narrow range of tilt angles, with the hard axis along $\bip$.
Similar observations were made in wide GaAs/AlGaAs quantum wells with two occupied subbands \cite{pan:2001,luhman:2006}.
However, we are not aware of any reports that $\bip$ can induce significant anisotropy near half-integer $\nu$ in a wide range of $N \geq 2$ in originally isotropic 2D systems.
 
In this Rapid Communication we report on a strongly anisotropic transport in a 2D hole gas (2DHG) in a high-mobility Ge/SiGe quantum well \cite{dobbie:2012}.
While no significant anisotropy is observed in either purely perpendicular or purely parallel $B$ (up to at least $B = 10$ T), tilted $B$ introduces a dramatic anisotropy.
Remarkably, the anisotropy emerges almost everywhere, except for QH states, with the hard (easy) axis oriented parallel (perpendicular) to $\bip$, and is largely  controlled by a single parameter, the tilt angle $\theta$, up to $N \sim 20$.
Although the emergence of the anisotropy naturally hints on a stripe phase, our findings differ from observations in GaAs in several important aspects, including the lack of significant anisotropy at $\bip =0$, easy tunability by $\theta$, and persistence to much higher $N$ and $T$.

%%%%%%%%%%%%%%%%%%%%%%%%%%%%%%%%%%%
\begin{figure}[b]
\vspace{-0.2 in}
\includegraphics[width=\linewidth]{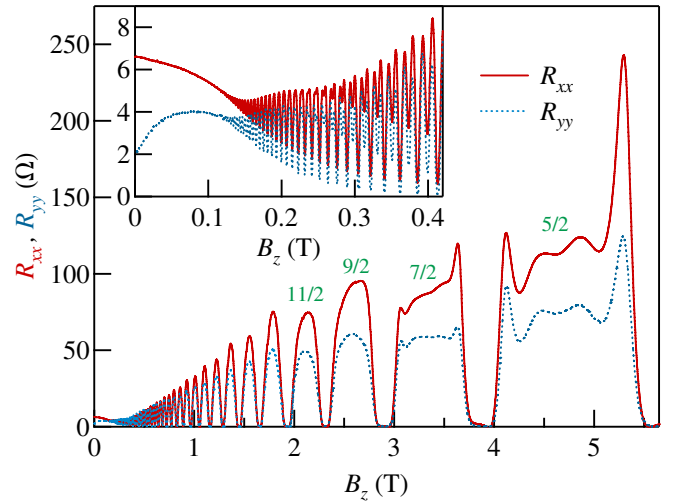}
\vspace{-0.25 in}
\caption{(Color online)
$\rxx$ (solid line) and $\ryy$ (dotted line) versus $B_z$ at $\theta = 0$ and $T = 0.3$ K. 
%Inset shows the data at low $B_z$.
}
%\vspace{-0.2 in}
\label{fig.1}
\end{figure}
%%%%%%%%%%%%%%%%%%%%%%%%%%%%%%%%%%

%%%%%%%%%%%%%%%%%%%%%%%%%%%%%%%%%%%
\begin{figure*}[t]
%\vspace{-0.2 in}
\includegraphics[width=\textwidth]{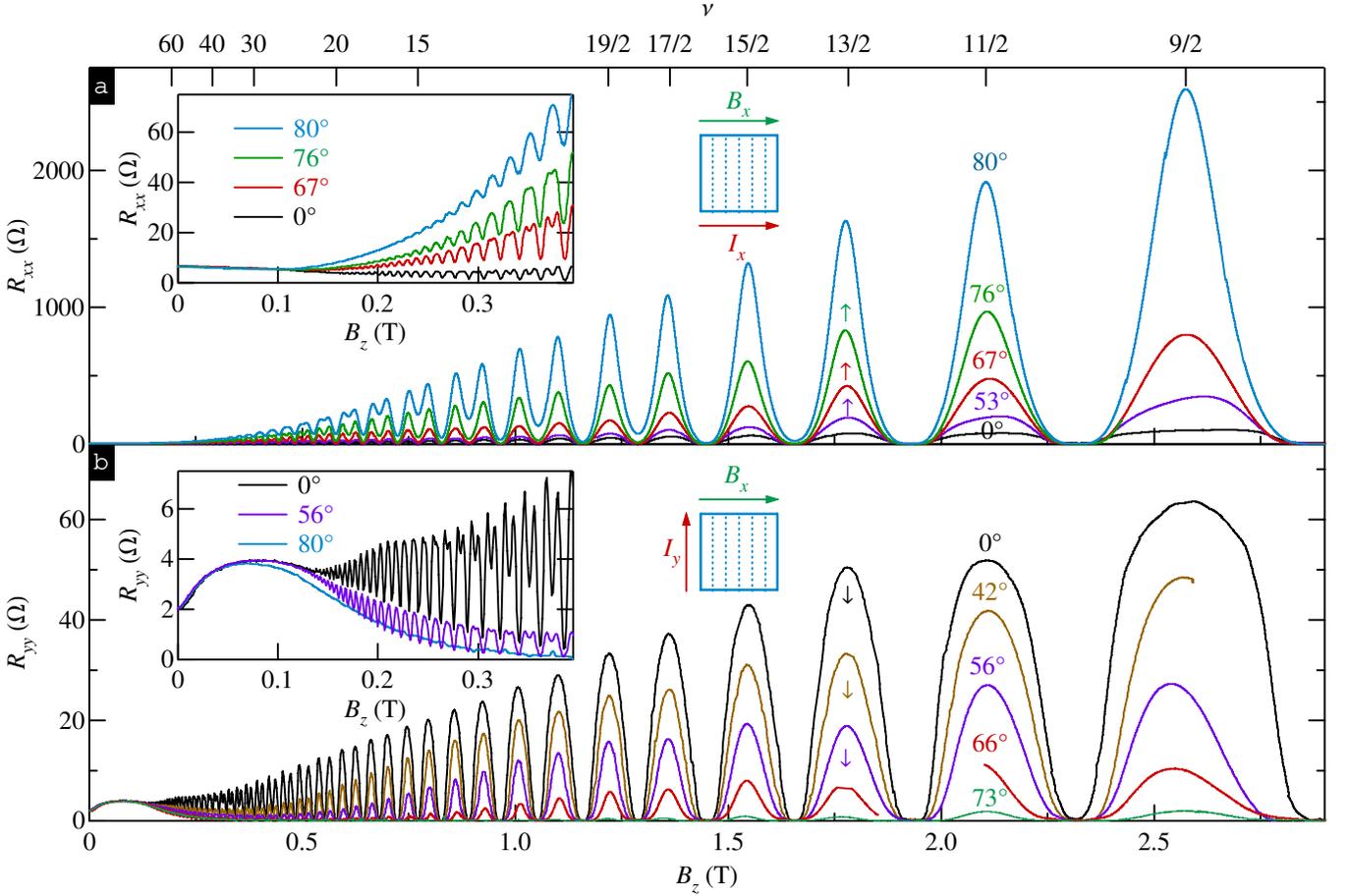}
\vspace{-0.25 in}
\caption{(Color online)
(a) $\rxx$ and (b) $\ryy$ at different $\theta$ versus $B_z$ (bottom) and $\nu$ (top) at $T \approx 0.3$ K. 
}
\vspace{-0.2 in}
\label{fig.2}
\end{figure*}
%%%%%%%%%%%%%%%%%%%%%%%%%%%%%%%%%%

Unless otherwise noted, the presented data were obtained on a $\sim5\times5$ mm square sample fabricated from a fully strained, 20 nm-wide, Ge quantum well grown by reduced pressure chemical vapor deposition on a relaxed Si$_{0.2}$Ge$_{0.8}$/Ge/Si(001) virtual substrate \citep{dobbie:2012,morrison:2014,myronov:2014,myronov:2015}.
At $T$ = 0.3 K, our 2DHG has density $p \approx 2.8 \times 10^{11}$ cm$^{-2}$ and mobility $\mu \approx 1.3 \times 10^6$ cm$^2$/Vs. 
The resistances $\rxx \equiv R_{\left < 1\bar{1}0 \right >}$ and $\ryy\equiv R_{\left <110 \right >}$ were measured by a low-frequency lock-in technique. 

Before presenting our results, we briefly discuss how our 2DHG in Ge/SiGe compares to 2D systems in GaAs/AlGaAs. 
First, Ge (GaAs) has a diamond (zinc blende) crystal structure which has (lacks) an inversion center. 
Second, the perpendicular component of the g-factor in Ge is much larger than in GaAs, while its parallel component is zero \cite{winkler:2003}, resulting in a much larger, but $\bip$-independent, Zeeman energy.
On the other hand, the band structure in our 2DHG is relatively simple; the light hole band is pushed down by strain and only the heavy hole band, with an effective mass $m^\star \approx 0.09\,m_e$ \cite{zudov:2014,shi:2014b,morrison:2014}, is populated.
In this respect, a 2DHG in Ge/SiGe is more akin to a 2DEG than to a 2DHG in GaAs/AlGaAs.

%fig.1
In \rfig{fig.1} we present $\rxx$ and $\ryy$ versus $B_z$ at $\theta = 0$ and $T = 0.3$ K. 
As shown in the inset, quantum oscillations corresponding to even (odd) $\nu$ start to develop at $B_z \approx 0.1$ T ($\approx 0.25$ T). 
At higher $B_z$, both $\rxx$ and $\ryy$ show QH states at all integer $\nu$, attesting to excellent quality of our 2DHG \citep{note:fqhe}.
While $\rxx$ and $\ryy$ differ by about a factor of three at $B_z = 0$, no strong anisotropy is observed at $B_z \gtrsim 0.1$ T. %\cite{note:0}. 
However, as we show next, once $\bip$ is introduced, a remarkably strong anisotropy sets in.

%%%%%%%%%%%%%%%%%%%%%%%%%%%%%%%%%%%
\begin{figure}[b]
\vspace{-0.2 in}
\includegraphics[width=\linewidth]{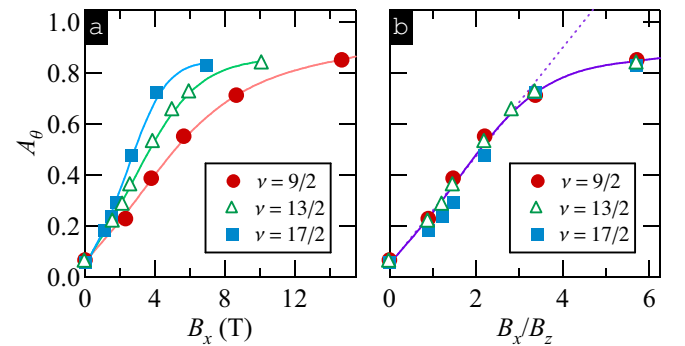}
\vspace{-0.2 in}
\caption{(Color online)
$\A$ versus (a) $B_x$ and (b) $B_x/B_z$ at $\nu= 9/2$, $13/2$, and $17/2$.
Solid lines are guides for the eyes.
Dotted line is drawn at $\A = 0.05 + 0.21 \tan \theta$.
}
%\vspace{-0.2 in}
\label{fig.3}
\end{figure}
%%%%%%%%%%%%%%%%%%%%%%%%%%%%%%%%%%%%

%fig.2
In \rfig{fig.2}(a) and \rfig{fig.2}(b) we present $\rxx$ and $\ryy$, respectively, versus $B_z$ (bottom) and $\nu$ (top), for different $\theta$ with $\bip=B_x$.
We observe that with increasing $\theta$, $\rxx$ ($\ryy$) increases (decreases) almost everywhere except at the QH states.
At $\nu = 9/2$ and $\theta = 80^\circ$ the resistance ratio reaches $\rxx/\ryy \simeq 3 \times 10^4$ ($\rxx \approx 2.6$ k$\Omega$, $\ryy < 0.1$ $\Omega$).
When $\bip = B_y$, the hard and easy axes switch places, i.e., $\rxx$ decreases and $\ryy$ increases, showing almost identical dependence on $\theta$.
Since the hard (easy) axis is always parallel (perpendicular) to $\bip$, the sole cause of the observed anisotropy is tilting the sample.
The intrinsic zero-field anisotropy, on the other hand, seems to be irrelevant. 

We define the anisotropy as
$\A \equiv (\rho_{xx}/\rho_{yy}-1)/(\rho_{xx}/\rho_{yy}+1)$, where $\rho_{xx}/\rho_{yy}$ 
is found using $(\pi\sqrt{\rho_{xx}/\rho_{yy}}/4 -\ln2) e^{\pi\sqrt{\rho_{xx}/\rho_{yy}}} \approx 4 \rxx/\ryy$ \cite{simon:1999}.
In \rfig{fig.3}(a) we present $\A$ versus $B_x$ for $\nu= 9/2$, 13/2, and 17/2.
We find that $\A$ starts at $\A \approx 0.05$, increases approximately linearly with $B_x$, and eventually saturates. 
We observe that at higher $\nu$, smaller $B_x$ is needed to induce the same $\A$.
Remarkably, the data at all $\nu$ can be well described by a common dependence on $B_x/B_z = \tan\theta$.
Indeed, as illustrated in \rfig{fig.3}(b), $\A$ versus $B_x/B_z$ for all $\nu$ fall onto a single curve.
Such a dependence is quite remarkable and we are not aware of similar findings in GaAs. 
The dotted line, drawn at $\A = 0.05 + 0.21 B_x/B_z$, illustrates that $\A$ increases roughly linearly until $B_x/B_z \approx 3$.

%%%%%%%%%%%%%%%%%%%%%%%%%%%%%
\begin{figure}[t]
%\vspace{-0.2 in}
\includegraphics[width=\linewidth]{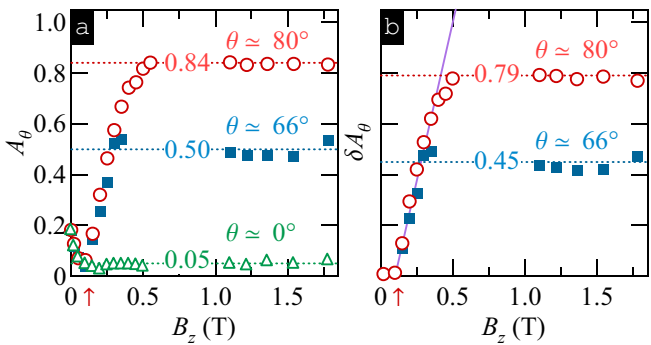}
\vspace{-0.25 in}
\caption{(Color online)
(a) $\A$ versus $B_z$ for $\theta \simeq 0^\circ$, $66^\circ$ and $88^\circ$.
(b) $\delta \A$ versus $B_z$ for $66^\circ$ and $88^\circ$.
Solid line represents $\delta \A = (B_z-B_0)/B_\star$, where $B_0 = 0.1$ T and $B_\star = 0.4$ T.
}
\vspace{-0.2 in}
\label{fig.4}
\end{figure}
%%%%%%%%%%%%%%%%%%%%%%%%%%%%%%%%%%%%

To see how $\A$ evolves with $B_z$ we construct \rfig{fig.4}(a) showing $\A(B_z)$, for $\theta = 80^\circ$, $66^\circ$, and $0^\circ$.
Below $0.1$ T, $\A$ is independent of $\theta$ and decreases with $B_z$.
At higher $B_z$,  $\A$ increases and saturates at $\approx 0.84\,(0.50)$ for $\theta = 80^\circ\,(66^\circ)$.
In \rfig{fig.4}(b) we present $\delta \A = \A - A_{\theta = 0^\circ}$ demonstrating that at $B_z \lesssim B_0 = 0.1$ T, $\bip$ does not induce any anisotropy.
A roughly linear growth of $\delta \A$ with $B_z$ follows $\delta \A = (B_z-B_0)/B_\star$, where $B_\star = 0.4$ T (cf. solid line).
The data at \emph{both} angles are described well by this dependence until $\delta \A$ saturates at  $B_z \approx B_{\theta} \approx 0.5$ $(0.3)$ T at $\theta = 80^\circ$ $(66^\circ)$.
We thus conclude that at $B_0 < B_z <B_\theta$, $\A$ is controlled primarily by $B_z$. 
At $B_z > B_\theta$, $\A$ is independent of both $B_z$ and $B_x$ for a given $\theta$, which again confirms that $\A$ is controlled by $\theta$ alone.
In contrast, the native anisotropy in GaAs increases with $B_z$ until it vanishes at $N <2$.

\begin{figure}[t]
%\vspace{-0.2 in}
\includegraphics[width=\linewidth]{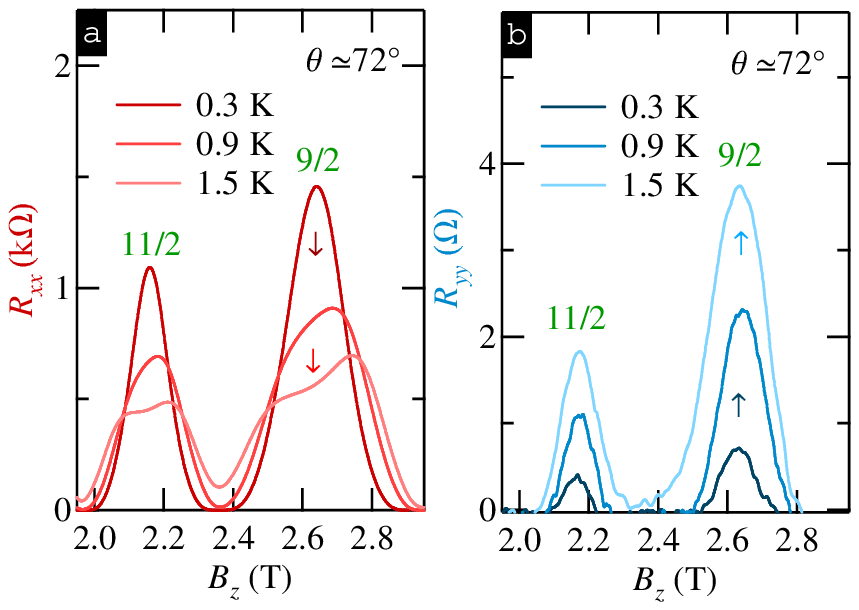}
\vspace{-0.25 in}
\caption{(Color online)
(a) $\rxx$ and (b) $\ryy$ (solid line) at $\theta \approx 72^{\circ}$ ($\bip = B_x$) and $T = 0.3$, 0.9 and 1.5 K. 
}
\vspace{-0.20 in}
\label{fig.5}
\end{figure}

We next demonstrate that the observed anisotropy is remarkably robust against temperature. 
\rfig{fig.5} shows (a) $\rxx$ and (b) $\ryy$ measured in different sample at $\theta \approx 72^{\circ}$ ($B_x/B_z \approx 3$) and $T = 0.3$, 0.9 and 1.5 K. 
At filling factor $\nu = 9/2$, the ratio $R_{xx}/R_{yy}$ exceeds $2000$ at $T = 0.3$ K and drops by about an order of magnitude as the temperature is raised to $T = 1.5$ K.
This drop occurs due to \emph{both} decreasing $\rxx$ \emph{and} increasing $\ryy$ (which change much more rapidly than in the isotropic state at $\theta = 0$), suggesting that the anisotropy will vanish completely at a few kelvin.
Interestingly, the $\rxx$ maxima at half-integer $\nu$ evolve into local minima with increasing $T$. 
%%%%%%%%%%%%%%%%%%%%%%%%%

%%%%%%%%%%%%%%%%%%%%%%%%%%%%%%%%%%%
%%%%%%%%%%%%%%%%%%%%%%%%%%%%%%%%%%

While we cannot currently explain why tilted field induces such a strong and robust anisotropy in Ge, below we examine several scenarios.
First obvious scenario is the formation of stripes, similar to those found in GaAs.
Indeed, as no significant anisotropy shows up in a purely perpendicular magnetic field (even at lower temperatures), an in-plane magnetic field is an essential ingredient for the observed anisotropy. 
We recall that the original prediction of the stripe phase \cite{koulakov:1996,fogler:1996} \emph{did not} specify any preferred direction in the 2D plane, i.e., it predicted randomly oriented stripe domains and no anisotropy on a macroscopic scale.
Thus, one possibility is that $\bip$ aligns these pre-existing stripe domains giving rise to the macroscopic transport anisotropy.
According to \rref{sodemann:2013}, the native anisotropy in GaAs results from a \emph{combination} of Rashba and Dresselhaus spin-orbit interactions.
Since Ge lacks the Dresselhaus term, such a symmetry-breaking mechanism does not apply and no native macroscopic anisotropy should be expected.
However, introducing an external field, such as $\bip$, could indeed reveal the underlying stripe phase producing observed anisotropy.
Furthermore, since $\bip$ is the only symmetry-breaking field in our 2DHG, one can indeed expect easy tunability and simple dependence on $\theta$, in contrast to complex behavior in GaAs caused by the interplay between $\bip$ and other symmetry-breaking fields.
We also note that the direction of the anisotropy axes with respect to $\bip$ is consistent with what has been observed in GaAs, especially at initially isotropic filling factors, such as $\nu$ = 5/2 and 7/2.

On the other hand, there exist factors which seem to rule out stripes as the origin of the anisotropy in our 2DHG, namely, the persistence to much higher $N$ and $T$ compared to that in GaAs.
Indeed, at such high temperature, no strong anisotropy has been observed in GaAs, even under applied $B_\parallel$.
Although $B_\parallel$ can change stripe orientation, theory predicts very small energy difference ($\sim 10^{-2}$ K) between stripes being parallel and perpendicular to $B_\parallel$ \cite{jungwirth:1999,stanescu:2000}. 
The persistence of the anisotropy in Ge up to $T > 1$ K suggests a much larger energy scale.
It would be interesting to test the possible existence of anisotropic domains in a purely perpendicular field.
For example, nuclear magnetic resonance \cite{friess:2014} and pinning mode resonances in the r.f. conductivity \cite{zhu:2009} are promising techniques to probe such domains.
Other external perturbations, such as direct current, in principle, could also align the domains and lead to macroscopic anisotropy \cite{gores:2007}.

It is known that $\bip$ couples the 2D cyclotron motion to the motion in the $\hat z$ direction due to finite thickness effects \cite{note:4}.
This coupling results in the anisotropy in both the effective mass \cite{khaetskii:1983,smrcka:1990,smrcka:1994,hatke:2012c} and in the Fermi contour \cite{kamburov:2012,kamburov:2013b}.
However, However, for $\bip = B_x$, this mechanism leads to $\rxx < \ryy$ which is
opposite \cite{dassarma:2000} to what we observe in our experiment.

Finally, we mention that surface roughness, in combination with $\bip$, was proposed \citep{chalker:2002} to explain anisotropies near level crossings \citep{zeitler:2001,pan:2001}.
However, such a scenario is not applicable here since our 2DHG is a single-band system and the vanishing in-plane component of the $g$-factor \cite{winkler:2003,drichko:2014} precludes crossings of spin sublevels. 
Although the surface roughness can lead to modest anisotropies in zero field \citep{hassan:2014} or in pure in-plane magnetic fields \cite{goran:2008}, it is not clear how it could be linked to the observed anisotropy in the QH regime.
Since experiments on Ge quantum wells with much lower mobilities have found no transport anisotropies in tilted $B$ \cite{drichko:2014}, mobility seems to be an important parameter. It is indeed highly desirable to perform measurements on various samples to investigate how the anisotropy depends on mobility, carrier density, strain, symmetry, and width of the quantum
well.

In summary, we observed a strong anisotropy in the quantum Hall regime of a 2DHG in a Ge/SiGe quantum well.
The anisotropy
(i)  emerges \emph{only} in tilted $B$ and can be easily tuned by $\theta$,
(ii) is characterized by $\rxx/\ryy$ which can be as high as $3 \times 10^4$,
(iii) persists to high $\nu$, and
(iv) requires neither extremely low $T$ nor extremely high mobility.
These features set the observed phenomenon apart from the anisotropic phases in GaAs/AlGaAs and, as such, point towards a novel mechanism of transport anisotropy, which, for some reason, is suppressed in GaAs. 
As a result, observation of a distinct type of strongly anisotropic transport in a system other than GaAs represents an important step towards overall understanding of electronic anisotropies.

\begin{acknowledgments}

We thank G. Jones, S. Hannas, T. Murphy, J. Park, A. Suslov, and D. Smirnov for technical assistance with experiments, and P. Baity for assistance with wire bonding.
We thank S. Das Sarma, L. Engel, Z. Jiang, S. Kivelson, K. von Klitzing, A. MacDonald, R. Nicholas, W. Pan, D. Polyakov, M. Shayegan, B. Shklovskii, R. Winkler and K. Yang for discussions. 
The work at Minnesota was funded by the US Department of Energy, Office of Basic Energy Sciences, under Grant No. ER 46640-SC0002567. 
Q.S. acknowledges support from the Allen M. Goldman fellowship.
The work at Warwick was supported by a EPSRC funded ``Spintronic device physics in Si/Ge Heterostructures" EP/J003263/1 and ``Platform Grant" EP/J001074/1 projects.
Experiments were performed at the National High Magnetic Field Laboratory, which is supported by NSF Cooperative Agreement No. DMR-0654118, by the State of Florida, and by the DOE.
\end{acknowledgments}


\begin{thebibliography}{49}
\bibitem[{not({\natexlab{a}})}]{note:fqhe}
\bibinfo{note}{Our 2DHG also reveals fully developed QH states at $\nu = 5/3$,
  4/3, and 2/3 which, to our knowledge, have not been reported in the
  literature. However, observation of fractional QH effect in Ge falls outside
  the scope of the present work and is left for future publication.}
\end{thebibliography}
\end{document}